\documentclass[jcp,twocolumn,groupedaddress
preprint,
superscriptaddress,
showpacs,preprintnumbers,
amsmath,amssymb
]{revtex4}
\usepackage{graphicx}
\usepackage{dcolumn}
\usepackage{bm}%
\usepackage[colorlinks=true,citecolor=blue]{hyperref}
\hypersetup{colorlinks=true,citecolor=blue,linkcolor=red,urlcolor=blue}

\usepackage{color}
\definecolor{Green}{RGB}{0,204,102}
\definecolor{Purple}{RGB}{102,0,255}
\definecolor{Blue}{RGB}{51,153,255}
\definecolor{Red}{RGB}{151,010,010}

\newcommand {\be}{\begin{equation}}
\newcommand {\ee}{\end{equation}}
\newcommand {\bea}{\begin{array}}
	
	\newcommand {\eea}{\end{array}}

\newcommand{\tqr}{\textquotedblright}
\newcommand{\tql}{\textquotedblleft}

\begin{document}
\title{Bonding properties of amorphous silicon and quantum confinement in the mixed phases of silicon nano slabs }
	
\author{Zahra Nourbakhsh}
\email{z.nourbakhsh@ipm.ir}
\affiliation{School of Nano Science, Institute for Research in Fundamental Sciences (IPM), Tehran 19395-5531, Iran}
\affiliation{Farhangian University, Isfahan 81498-13518, Iran}

\author{Hadi Akbarzadeh}
\affiliation{Department of Physics, Isfahan University of Technology, 84156-83111 Isfahan, Iran}

\begin{abstract}
On the basis of density functional calculations and using Bader's atom in molecule theory, this article presents quantitative microscopic analyses on the bonding properties of amorphous silicon (a-Si) which could reflect in the observable mechanical and electronic behaviors of this material. In addition, the occurrence and strength of quantum confinement of charge carriers in a composition of silicon crystal nano slabs (SiNSs) embedded in hydrogenated a-Si (a-Si:H) semiconductor are studied.  It is shown that the strongest confinement effect happens for Si slabs limited in [100] direction.  The band gap tunability with the width of SiNSs is exhibited and a scaling law is investigated for the size dependent behavior of energy states. It is demonstrated and argued why in these systems the confinement of holes is stronger than electron confinement. The computational methodology used to passivate a-Si defects by hydrogen is also detailed.
\end{abstract}
	
\maketitle	

\section{Introduction}
\label{Introduction} 

Sunlight is the most accessible source of renewable energy, nevertheless the practical use of solar energy is still very low due to the high cost of photovoltaic energy \cite{qc-app}. The third generation solar cell systems, by using quantum structures, try to increase the efficiency and decrease the costs. In this regard, silicon plays the main role in the optoelectronic industry owing to its band gap energy which is in the favorable region of the solar radiation spectrum \cite{amoSi, shq}. Additionally, silicon is the second most common element in the earth's crust.

Confinement of charge carrier wave functions in the nano structures provides a means of controlling and engineering the absorption band gaps of nanocrystalline materials~\cite{qc}. It also increases the efficiency of multiple exciton generation \cite{meg} and enhances charge carrier life time which is promising for ultimately improving the solar cell efficiency \cite{qc-app}.
The confinement power of vacuum is stronger than any other dielectric matrices; however, the mobility and conductivity of charge carriers between free-standing nano particles (NPs) are low. Encapsulation of NPs in a semiconductor with a wider band gap could produce advantages of both confinement and charge transfer.  For example QC was found in Si nanocrystals embedded in Si-oxides \cite{oxide}, Si-nitrides \cite{nitride} and also amorphous-Si (a-Si) \cite{mattoni, lusk1, lusk2}. The matrix geometry also provides the opportunity of building tandem solar cells and going beyond the standard single junction solar cells efficiency \cite{tandem}. 

The probable confinement of charge carriers in the multi-phase Si system 
with only topological changes in barriers opens new doors for research in designing and engineering novel configurations. This mixed phase system is a composite totalizing the advantage of two separate phases in one material.
Since the energy gap in diamond silicon (d-Si) crystal (1.1~eV \cite{mattoni}) is less than (hydrogenated) a-Si (1.5-1.9~eV \cite{mattoni}), QC in d-Si NPs embedded in a-Si:H matrix, named nc-Si:H, could provide the energy gap ranging in the high intensity area of solar energy radiation \cite{shq}.  Additionally, the close alignment between energy states of Si nanocrystals and a-Si phase increases the tunneling probability and  conductivity of charge carriers and reduces the rate of light induced degradation (LID) \cite{lid}. Besides, this is an environment-friendly photovoltaic product.

There are several theoretical and experimental researches on nc-Si:H system \cite{mattoni, mattoni1, mattoni2, mattoni3, lusk1, lusk2} which report different interesting features of QC phenomena in this system. This system exhibits a novel kind of QC happening just for the valence band states (holes) \cite{mattoni}; additionally, it is shown that in the absence of hydrogenation, confinement is not possible even in the presence of four fold coordination in a-Si \cite{mattoni1}. 
Besides the non equilibrium thermodynamic and kinetic effects on the micro structure and the hydrogenation, like clustering phenomena occurring during crystallization, are discussed in Ref. \cite{mattoni2}.  
However, all aspects of this system have not been known yet. The focus of this article is on the role of a-Si matrix.

a-Si is one of the most promising photovoltaic semiconductors; it has higher flexibility but lower electronic efficiency in comparison with d-Si crystal. The defects in a-Si cause appearing some undesirable electronic states which lead to reduce the confinement power of this matrix. In the hydrogenated a-Si (a-Si:H), the Si atoms inducing these defects are removed and the dangling bonds are passivated by H atoms.
This process is experimentally doing by introducing H during the a-Si fabrication. 

Here, on the basis of density functional Kohn-Sham theory \cite{k-sh} and using Bader's \tql atoms in molecules\tqr~ quantum theory \cite{bader}, we are interested to evaluate for the first time the distribution of the fine properties of chemical bonds of a-Si covalent material. Since the physical properties of a system strongly depend on its microscopic features, such a fundamental study could provide useful tools to explore and characterize other physical aspects of a-Si like its mechanical and electronic properties. 
The Bader's analyzing has been successfully used to study or illustrate many condensed matter phenomena; such as magnetic, structural or crystalline-amorphous phase transitions \cite{bader1, direction}, band structure properties \cite{nour}, dynamical softening and structural stability \cite{bulkm1, bulkm2, bader3}, exploring the effects of vacancies, defects, and doping in crystalline and amorphous materials \cite{bader4}, etc. 

\begin{figure}
\centering
\includegraphics[width=0.95\linewidth]{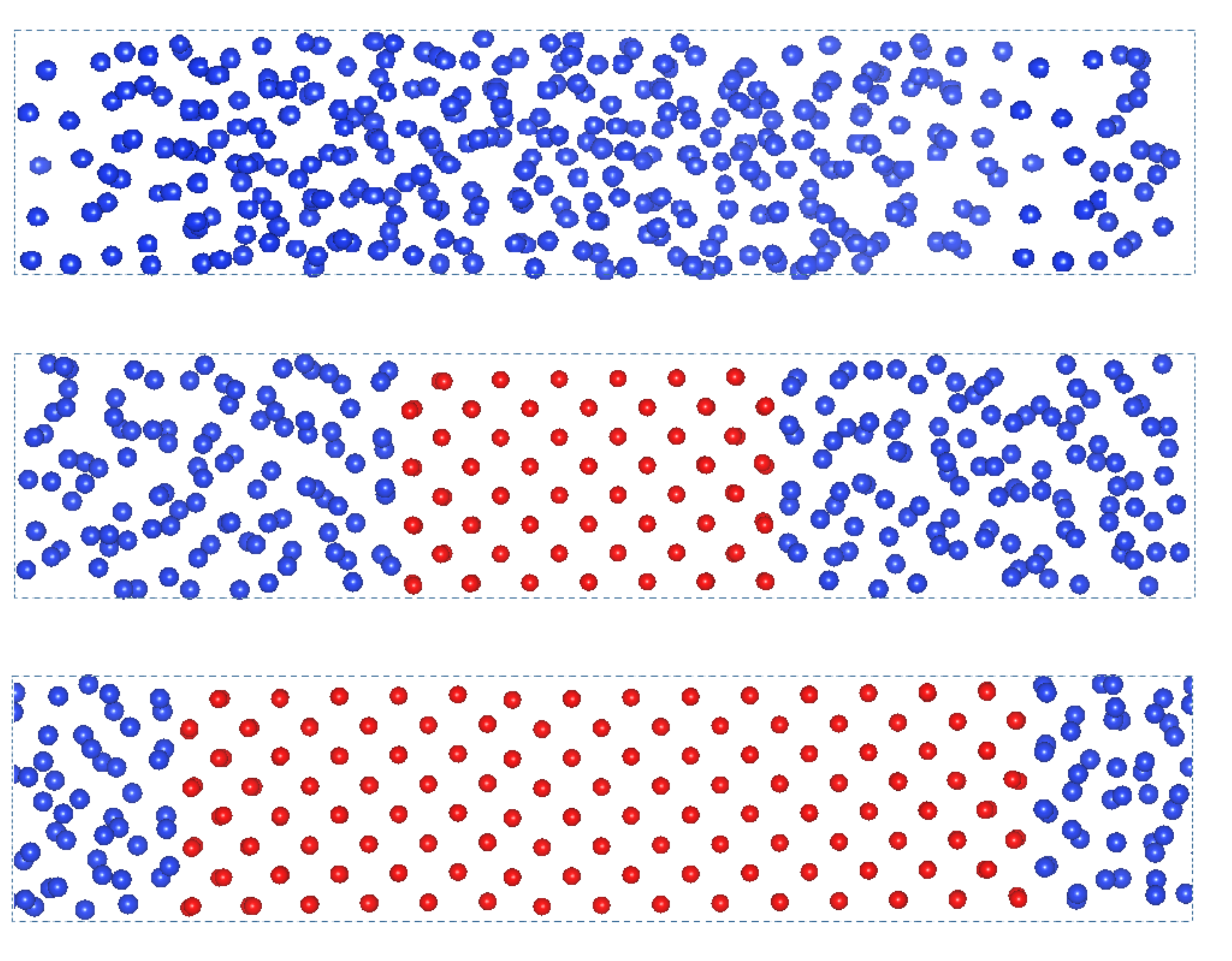}
\caption{\label{struct} Three of the six studied structures; SiNSs in [100] direction have been encapsulated within a-Si matrix of the same density. There are 320 silicon atoms in the mixed phases of crystalline and amorphous Si. The widths of SiNS (a-Si) of the considered systems are 0~(5.45), 0.54~(4.91), 1.63~(3.82), 2.72~(2.73), 4.09~(1.36) and 5.45~(0)~nm; the crystalline part is composed of rectangular grid of $2\times2\times n$ conventional unit cells ($n$ = 0, 1, 3, 5, 7, and 10). }
\end{figure}

Additionally, the second part of the paper presents the effect of the quality of the a-Si matrix on confinement of charge carriers in d-Si nano slabs (SiNSs) embedded within an a-Si:H matrix. 
This section also discusses on the directional effects on the strength of QC in nc-Si:H system.
Figure~\ref{struct} shows three of the six studied structures; the total number of atoms are conserved in all structures, so the amorphous size increases as crystal size decreasing. In Sec.~\ref{sec:qc}, we will show that the size reduction of a-Si matrix does not affect on the results. The slab geometry with the size independent planar boundaries is an appropriate system to clear the role of a-Si matrix in confinement.

This work is organized in the following way. After an overview of the computational approach and explanation of the hydrogenation method in the next section, the results are presented in Sec.~\ref{sec:result}. Analyzing and description of the bond properties of the amorphous system, exploring the directional effects on QC in free-standing SiNSs, studying the effect of a-Si matrix on QC, and explaining why QC happens just for one type of carriers in this system are discussed in this section. Finally Sec. \ref{sec:summary} gives a conclusion and a brief summary of the main results.

\section{Methodology} \label{sec:method}

In order to produce the realistic high-quality a-Si structures, the so called Wooten-Winer-Weaire (WWW) bond switching algorithm \cite{www, cleangap} is used; it conserves four-fold coordination by producing a continuous random network. The statistical properties of the generated a-Si structures, using this method, are in good agreement with experimental results  with standard deviations in bond lengths and bond angles as small as 0.073~\AA~ and $7^\circ$, respectively \cite{cleangap, exp}. The energy in amorphous phase is $\sim$ 0.14~eV/atom higher than that of d-Si crystal.  
 
The first-principles simulations based on density functional theory (DFT) are carried out using the SIESTA code \cite{siesta}. We use the generalized gradient approximation (GGA) of Perdew, Burke, and Ernzerhof (PBE) \cite{pbe}, along with norm-conserving pseudopotentials. A $double-\zeta$ polarized basis set including $3d$ and $4f$ orbitals for Si with an energy  mesh cutoff of 200 Ry is used. The geometrical optimization is performed and satisfied when all components of all forces are less than 0.04~eV/\AA. The studied systems have a supercell dimension of $10 \times 2 \times 2$ conventional unit cells (with the length of $5.45 \times 1.09 \times 1.09$~nm). 
In the calculations, periodic boundary conditions and only a single k-point are considered.
Because of using two conventional cells and periodic boundary conditions, the X-symmetry point (the conduction band minimum position in d-Si) is projected at Gamma point; the analysis of d-Si crystal shows that the associated gap is converged within 0.02~eV \cite{lusk1, lusk2}. 
 
In addition, several post-processing and home-made programs \cite{aim} are employed to describe and extract different fine structures of bonding properties of system on the basis of Bader's \tql atoms in molecules\tqr~ theory. According to this theory, the charge density critical points (CPs) are defined as the zeros of $\nabla \rho(r)$ where $\rho (r)$ is electron charge density at $r$ point.  The eigenvalues of the Hessian matrix of the electron charge density at the CPs ($|\nabla^2 \rho(r) - \lambda|=0$) are quantitative tools to analyze the CPs. According to the sign of $\lambda$s, there are four kinds of CPs for a three-dimensional field: a local maximum, a local minimum, and two kinds of saddle points.  A saddle point with only one of the three curvatures positive, is known as a bond CP (BCP). BCPs are usually placed on the line connecting two neighbor atoms;  the charge density of BCP is minimum along this direction ($\lambda_{\shortparallel}>0$) and maximum in the two perpendicular directions ($\lambda_{\perp}, \lambda'_{\perp}<0$). The bond ellipticity, which is determined as  $\frac{\lambda_{\perp}}{\lambda'_{\perp}}-1$, shows  electron  delocalization or electron anisotropy  of bond,  so higher value of ellipticity relates to weaker character of the bond. Another useful index, which can be used as a measure of bond instability is  directionality which is related to $\frac{|\lambda_{\perp}|}{\lambda_{\shortparallel}}$ or $\frac{|\lambda'_{\perp}|}{\lambda_{\shortparallel}}$ ratios \cite{direction};  and introduce the concept of closeness 
to bond breaking. Bond instability happens when $\lambda_{\perp} or \lambda'_{\perp}\rightarrow 0$, so higher value of directionality is proportional to more phase stability. Also $\Lambda=\sum_{i=1}^3 \lambda_i$ and the charge density at the bond points ($\rho _b$) are other important parameters to estimate the bond strength; the lower value of $\Lambda$ and higher value of the charge density of bond points refer to the stronger bond characteristic.  In what follows, we use these parameters to analyze bonding properties of a-Si system.
 
\paragraph*{\underline{a-Si Hydrogenation Method}} \label{sec:asi}

Due to the disorder nature of a-Si, some localized states appear in the valence and conduction band tails of this system. In order to enhance the energy gap and electronic properties of a-Si, the Si atoms producing these localized states should be removed and then the dangling bonds are passivated by hydrogen. 
To reach the optimal experimental value of hydrogen solubility in a-Si ($\sim 10\%$ \cite{hsol}), about $2\%$ of Si atoms are removed. So detecting the best Si atoms for removing is important.
To identify the defect states in a-Si, the method proposed by Allan $et~al.$ \cite{allan,lusk1,lusk2} is used but here the track is improved.
For this purpose, firstly, we select the removal states located in the gap area based on Allan formula; these states are distributed in less than $10-15\%$ of the total number of atoms. Then the Si atoms whose their orbitals contribute at the selected states are identified; but unlike the previous studies \cite{lusk1,lusk2}, all the removal atoms are not deleted in one step;
because it is possible that geometry optimization  stabilize the bonding at the neighboring area and the localization appears in the new part of system.

\begin{figure}
\centering
\includegraphics[width=0.95\linewidth]{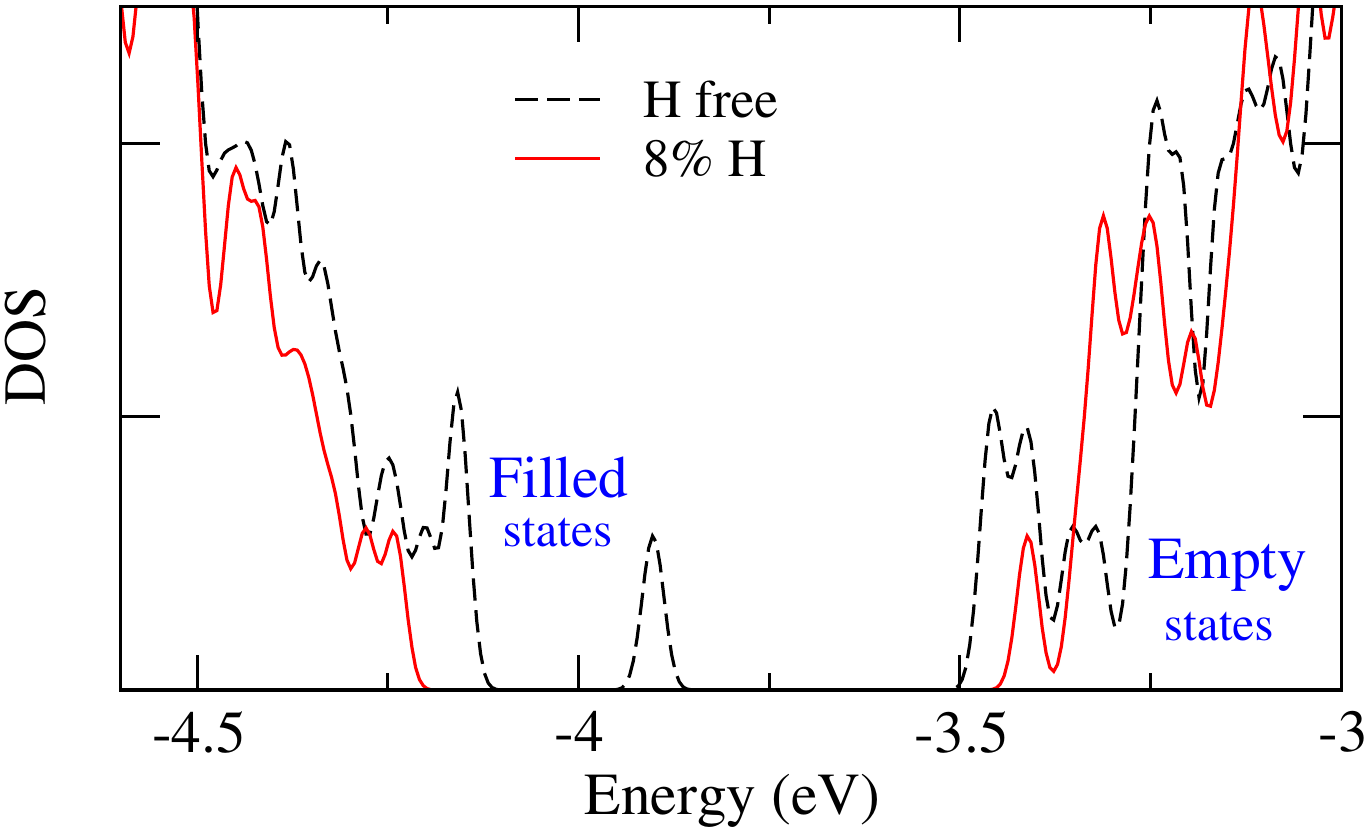}
\caption{\label{dos} The density of states of a 1000-atom a-Si system around the Fermi surface. The dashed black (solid red) line shows the DOS before (after) 8\% hydrogenation. }
\end{figure}

The hydrogenation of the a-Si system is done by a code developed by the authors which is implemented in SIESTA package.
Following this scheme, the a-Si energy gap increases $\sim$ 45-50~meV per H percent. 
This value is in good agreement with the TB results \cite{allan} but it is larger than experiment (12.7~meV per H percent \cite{expgap}). However by developing the experimental methods, the agreement could be increased.  Figure~\ref{dos} shows the impact of hydrogenation on the states around the band gap of a cubic a-Si supercell containing $5 \times 5 \times 5$ unit cells with 1000 Si atoms. 
 
In each of the studied structures, $\sim 2\%$ of Si atoms belonging to the amorphous part will be removed. In the following, we will show that hydrogen passivation restore the clean energy gap to a-Si which is $\sim$~0.35-0.40~eV higher than d-Si band gap and enables QC of carriers in nc-Si:H systems.

\section{Results}\label{sec:result}
 
\subsection{Bonding properties of a-Si }\label{sec:bond}
 
\begin{figure*}
\centering
\includegraphics[width=0.95\linewidth]{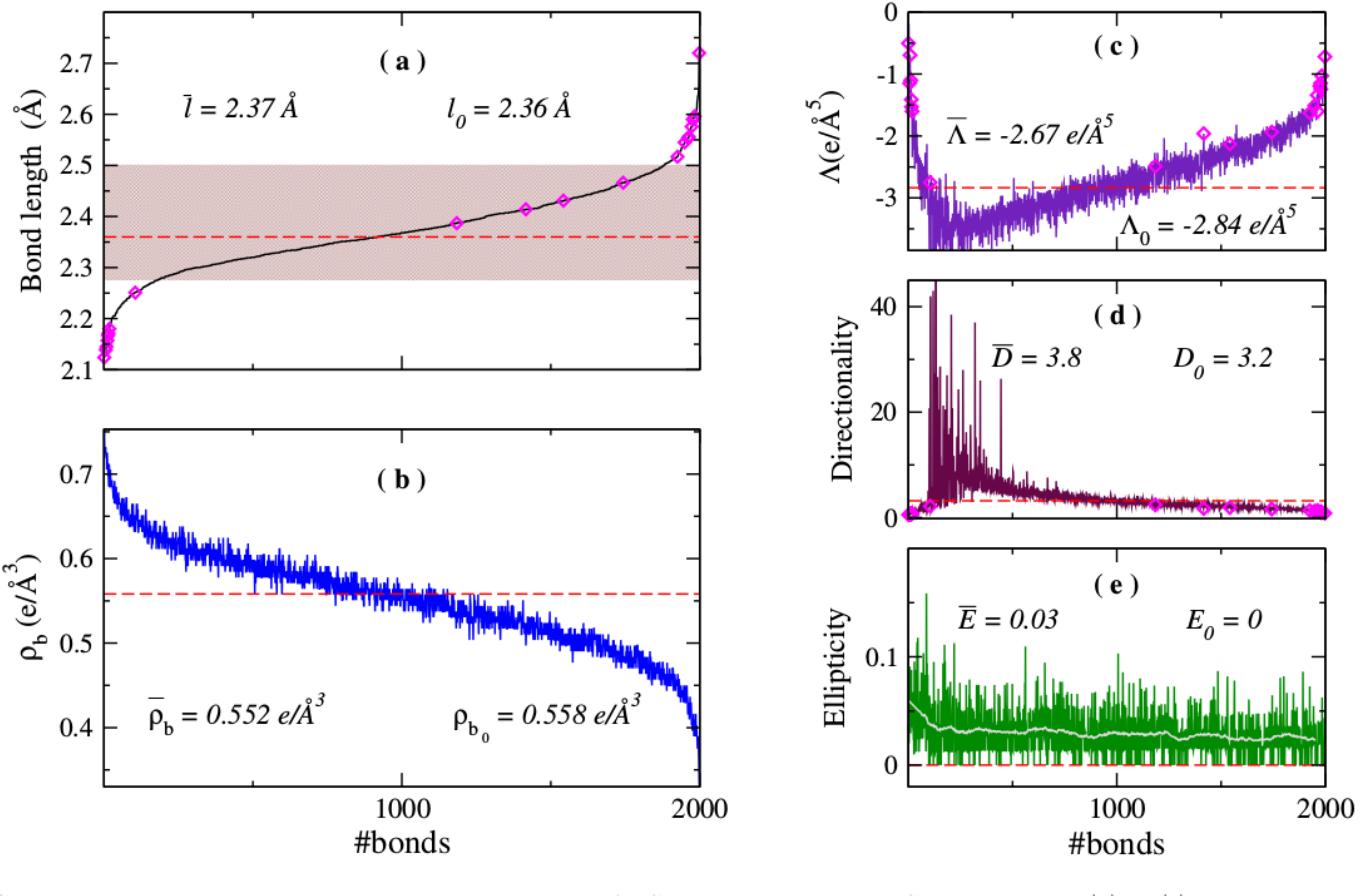}
\caption{\label{aSibond} The calculated equilibrium  bonding properties of a-Si structure with 1000 Si atoms. From (a) to (e): bond length, the charge density at the bond point, $\Lambda=\lambda_{\shortparallel}+\lambda_{\perp}+\lambda'_{\perp}$, directionality ($\frac{|\lambda_{\perp}|}{\lambda_{\shortparallel}}$ or $\frac{|\lambda'_{\perp}|}{\lambda_{\shortparallel}}$), and bond ellipticity. The red dashed lines show the corresponding value of d-Si. The diamond marks are related to the worst or weakest bonds of each removed silicon atoms. The white line in (e) shows the average value of ellipticity. All bond quantities have been sorted versus bond length. The average value of each diagrams as well as the corresponding value in d-Si crystal are also reported. }
\end{figure*}

This section begins with a quantum description of the a-Si chemical bonds. 
Due to the absence of periodicity, studying amorphous materials both experimentally and theoretically are more complicated than crystals. Here, we consider a cubic cell of a-Si with the length of $\sim$ 3~nm, containing 1000 Si atoms with the density same as d-Si crystal. This length is large enough to simulate an isotropic amorphous system. Periodic boundary conditions are applied in three extended directions after this length. In silicon with fourfold coordination, there are two bonds per Si atom, therefore 2000 covalent bonds exist in the 1000-atom a-Si system. The results summarized in Fig.~\ref{aSibond} determine the fine structures of all these bonds individually.

Different bond indexes of a-Si system are presented in Fig.~\ref{aSibond}; all diagrams in this figure are sorted by the bond length and displayed versus bond numbers. The dashed red lines in this figure display the corresponding quantities of d-Si crystal. Figure~\ref{aSibond}(a) shows the bond length distribution in a-Si material.  The shaded area in this figure separates the tails with exponential  distributions from the main part; about 85-90\% of bonds are in the shaded area around the bond length of d-Si. The difference between the smallest and the longest bonds in this area is $\sim$~0.2~\AA. 
The results given in Fig.~\ref{aSibond}(b) indicate that the charge density at the bond points decreases as a consequence of bond length enhancement. $\rho_b$ is a measure of covalent bond stiffness \cite{bulkm1, bulkm2}, which varies over the range 0.32-0.75~e/\AA$^3$. 
As noticed in the earlier section, the eigenvalues of the Hessian matrix of the electron charge density at the BCPs, namely $\lambda$s, provide further fine details of the system bonding properties. The effect of angle distortion in a-Si also reflects in $\lambda$s and their related parameters.
Figure~\ref{aSibond}(c) shows the $\Lambda=\sum_{i=1}^3 \lambda_i$ parameter for a-Si bonds, which is a negative quantity for covalent bonds and its smaller value expresses the stronger character of covalent bonding. Regards to Fig.~\ref{aSibond}(c), it is seen that the bonds with the small and long lengths show the weakest stability.
This behavior is confirmed by the bond directionality calculated in Fig.~\ref{aSibond}(d) which indicates that in a-Si, bonds with small Si-Si lengths are as brittleness as bonds with long lengths. This is an interesting finding, which comes from the stronger effect of bond distortion on bonds with smaller lengths. Therefore, despite the higher stiffness, these bonds are very brittle. $\Lambda$ and directionality behaviors illustrate that in the sequence of N$_b>$150, bond strength totally decreases by the bond length enhancement; however, there are some oscillations in these diagrams.
The bond ellipticity, presented in Fig.~\ref{aSibond}(e), is a measure of the anisotropic behavior of bond which does not follow a regular trend; however, its average value, shown with the white solid line in the figure, reveals that it decreases as a consequence of bond length enhancement which is a conclusion of the stronger bond distortion impacts on bonds with smaller length. The bond ellipticity of d-Si crystal is equal to zero.

The results presented in Fig.~\ref{aSibond}(a) indicate that $\sim$10-15\% of covalent bonds in a-Si material are in the unshaded area with a high brittleness. It means that up to 50\% (4$\times$12.5\%) of silicon atoms could have at least one brittle bond. 
As mentioned in the earlier section, we remove 2\% of Si atoms to promote the a-Si electronic properties. The diamond marks in Fig.~\ref{aSibond} show the most brittle bond of each removed Si atoms which are selected based on the method explained in Sec.~\ref{sec:method}. A strong correlation between weak bond characteristic and localized energy states in the gap position of a-Si is detected, however, the results show that several removed Si atoms are in the shaded area. Since the bond properties are almost a local behavior, it may reflect the effect of long range interactions in appearance of the localized states.

\begin{figure}
\centering
\includegraphics[width=0.95\linewidth]{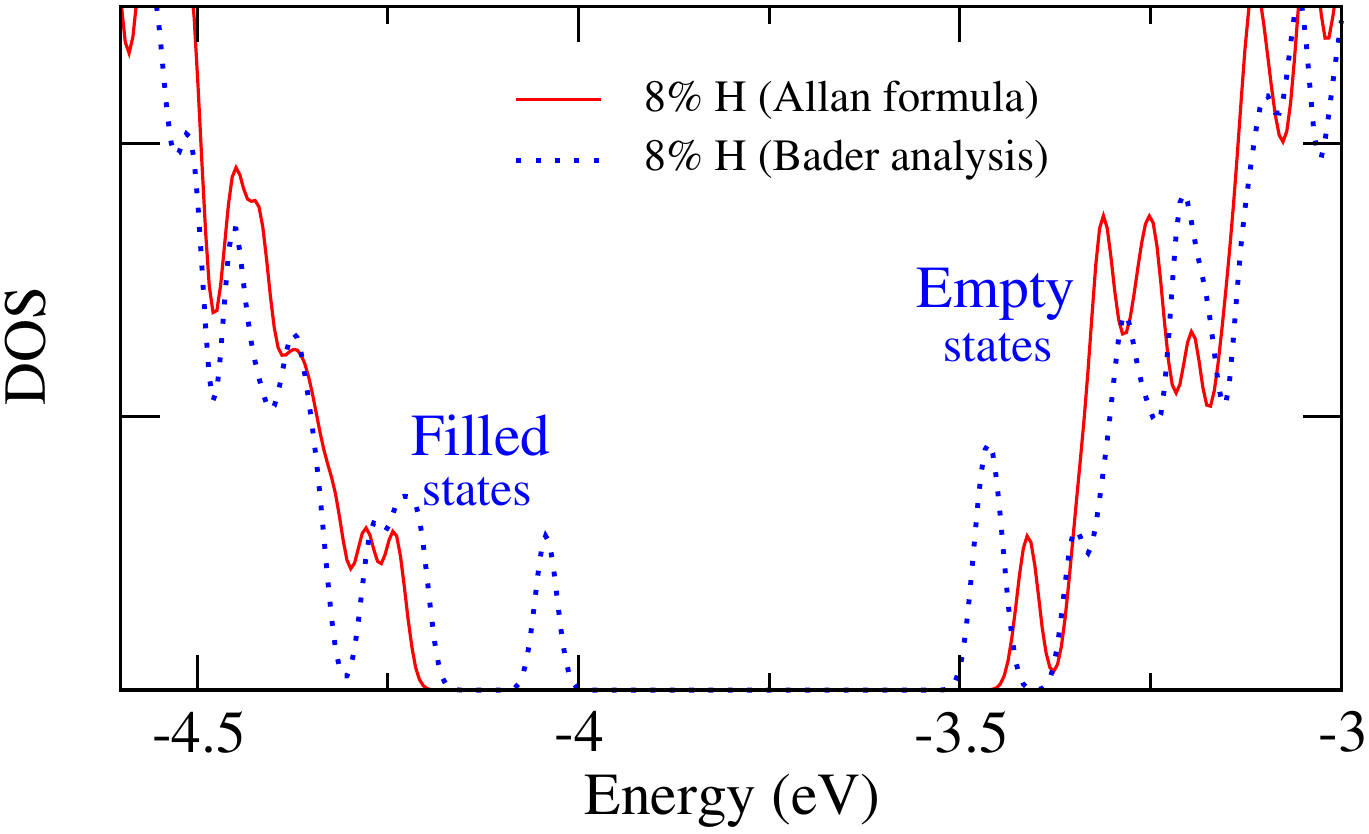}
\caption{\label{dos2} The density of states of the 8\% hydrogenated a-Si system with 1000-Si atoms. The solid red line represents the DOS of a-Si:H system displayed in Fig.~\ref{dos}; the removed Si atoms in this system are selected among the Si atoms which have localized states in the gap area based on Allan formula. The dotted blue line gives the DOS of a-Si:H system where 2\% of Si atoms with the highest brittle bonds are removed. }
\end{figure}

The question that arises here is what happens if all the removed Si atoms are those with most brittle bonds; does it lead to a  better electronic performance device or not? 
To answer this question, we have also studied the electronic structure of the 1000-atom a-Si system  when all the removed Si atoms are selected from the atoms with most brittle bonds.
Figure~\ref{dos2} compares the density of states of the a-Si:H system  introduced in the previous section with the DOS of new hydrogenated system. The results indicate that in the latter system, the energy gap increases just $\sim$~0.15~eV or 19~meV per H percent, which is strongly lower than the gap evolution in the a-Si:H system introduced in Sec.~\ref{sec:method}. Additionally, the hydrogenated system based on Allan formula is energetically more favorable. 

The impacts of defects in a-Si semiconductor have been mostly studied on its electronic structure while the mechanical properties of this system are not well known. Since the elastic constants are related to the bonding properties of the system, this study provides a novel description to estimate the mechanical properties of the system. The results presented in Fig.~\ref{aSibond} could show how percentage of bonds move toward stronger or weaker bonding. The average values of all quantities presented in Fig.~\ref{aSibond} as well as the corresponding values of d-Si are calculated and reported separately for each diagram in the figure. The results demonstrate that the average amounts are very close to those of d-Si values; this finding qualitatively agrees with the data available in the literature which predict that the bulk modulus of a-Si is very close to that of d-Si crystal \cite{aSiMech,aSiMech2}.
 
\subsection{QC in SiNSs : directional and a-Si effects }\label{sec:qc}

\begin{figure}
\centering
\includegraphics[width=0.95\linewidth]{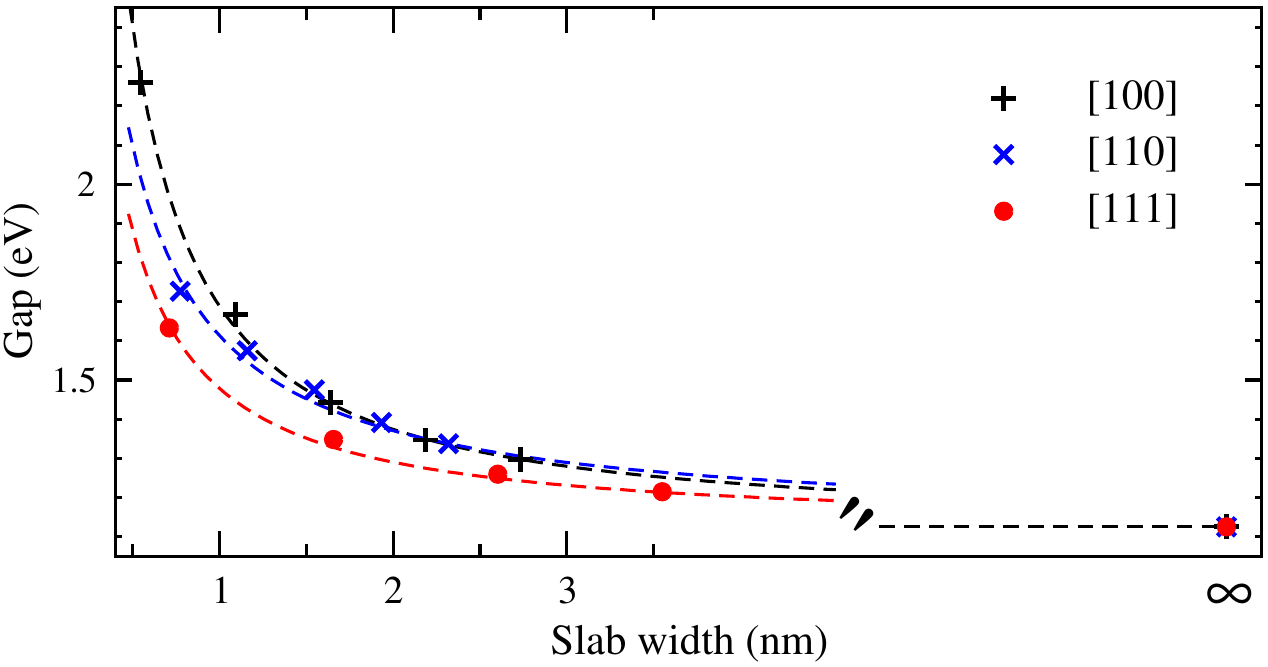}
\caption{\label{slabgap}  The evolution of the band gap as a function of the slab width in [100], [110], and [111] directions. All energies are shifted upward to set the bulk band gap of d-Si to be 1.12~eV.}
\end{figure}

In order to study the QC of charge carriers in SiNSs, we firstly check that the confinement effect for which direction of nano slabs is stronger.
The QC happens when the system size, at least in one direction, becomes less than exciton Bohr radius of the bulk system. This radius is in proportion to the inverse of the reduced exciton effective mass, $a^*_B \propto \frac{1}{\mu_{ex}} = \frac{m_e + m_h}{m_e \times m_h}$ \cite{grosso}. Regarding to the silicon band structure, it is not easy to estimate that the average effective mass in which direction is stronger. Consequently 
to this aim, Fig.~\ref{slabgap} exhibits the gap evolution for SiNSs, cut in different [100], [110] and [111] directions, versus slab width. All the surface dangling bonds are passivated by H atoms. 
The results reveal that QC is stronger (weaker) in [100] ( [111] ) direction. Thus in what follows, the impact of amorphous matrix on SiNSs cut in [100] direction will be studied.

\begin{figure}
\centering
\includegraphics[width=0.95\linewidth]{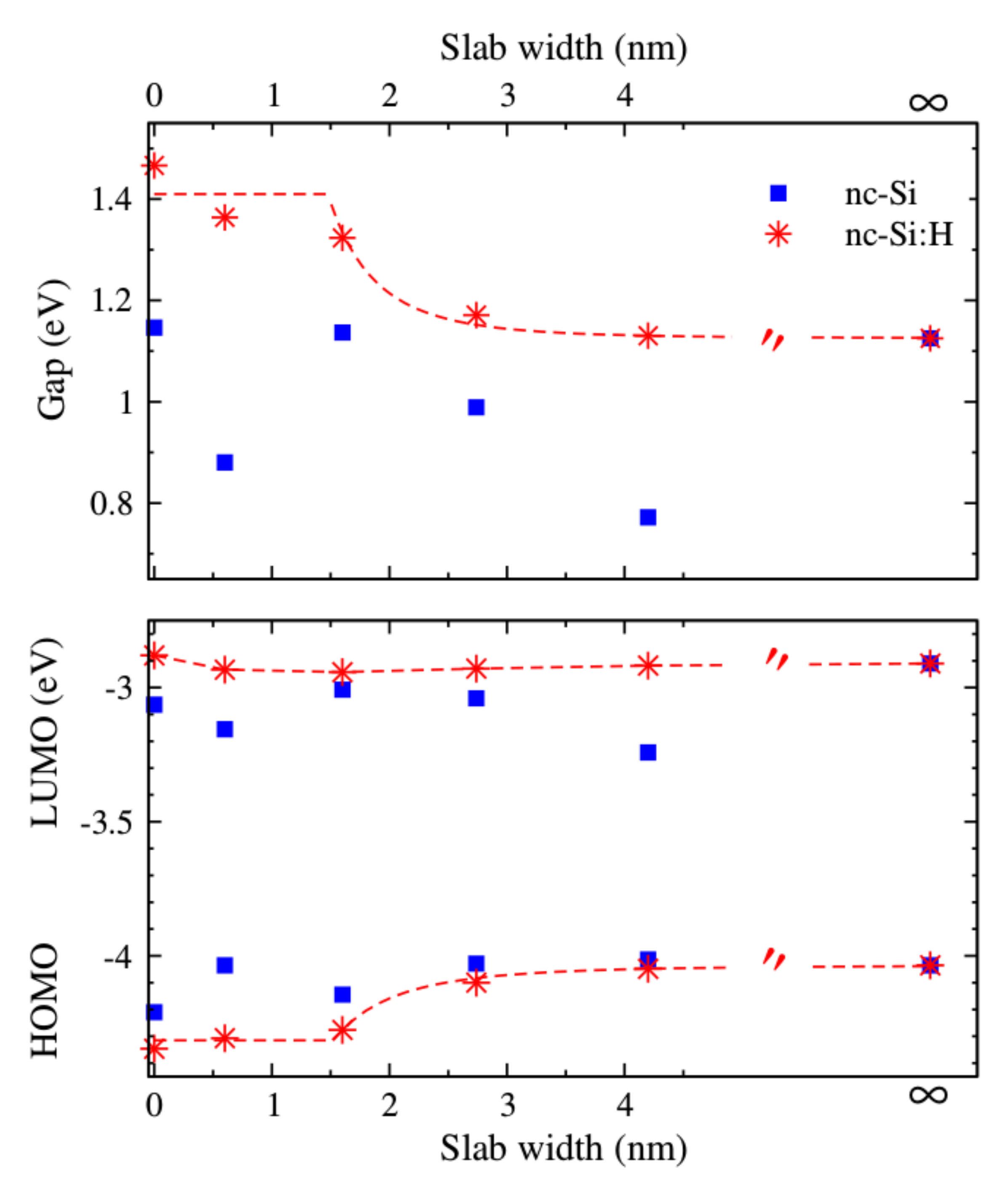}
\caption{\label{gap}  
Calculated energy gap, the HOMO, and the LUMO energy states as a function of slab width in the crystalline/amorphous composites presented in Fig.~\ref{struct}; the squares (stars) show these states before (after) hydrogenation. The dashed lines show power law fitting of the energy evolutions vs size. The LUMO energy states are shifted upward to set the d-Si band gap to be 1.12~eV. }
\end{figure}

Because of the band gap underestimation in DFT calculations \cite{dftgap3}, the gap energy of d-Si system is set to be 1.12~eV, which is the experimental band gap of silicon crystal, and LUMO and gap energies in Figs.~\ref{slabgap} and \ref{gap} are shifted upward by $\sim$ 0.5~eV. It is worth mentioning that the focus of this study is to reveal the general trends of system gap modulation by the charge carriers confinement; and shifting the energies by a constant value does not affect the main conclusions.

Now, we are ready to study QC in nc-Si:H mixed phases in the structures shown in Fig.~\ref{struct}. Since having just a single $k$-point, the molecular terminologies are used to describe the energy states.  The calculated HOMO (highest occupied molecular orbital), LUMO (lowest unoccupied molecular orbital), and gap which is the difference between LUMO and HOMO energies, are plotted versus width of SiNSs in Fig.~\ref{gap} for both nc-Si and nc-Si:H systems. The results indicate that confinement does not occur in the pure silicon systems while in the hydrogenated system, decreasing the crystalline size is accompanied with increasing the energy gap from the bulk value to the a-Si gap.

Regarding Fig.~\ref{gap}, for the smallest encapsulated SiNS, the strong interface effects prevent the confinement of carriers. For the larger systems, as plotted in the figure, the system energy gap is fitted with a power law function of the form $A/w^{\alpha}+E_{w\rightarrow\infty}$ where $w$ is the width of SiNSs and $E_{w\rightarrow\infty}$ is d-Si band gap. The confinement sensitivity coefficient (A) and the confinement power ($\alpha$) have been respectively found $A= 1.26$ and $\alpha = 3.8$. Additionally, figure~\ref{gap} predicts QC in the SiNSs for width larger than the critical width of $d_c \sim$~1.5~nm. The gap evolution in this system  is $\sim$~0.3~eV which is slightly smaller than that of the stand-alone SiNSs displayed in Fig.~\ref{slabgap} for the same range of slab widths, this difference refers to stronger confinement power of vacuum in comparing with a-Si:H. 
In the previous studies on nc-Si:H \cite{lusk1,lusk2}, the confinement sensitivity coefficient was obtained very small, $A \sim 0.4$, and consequently, the gap evolution was smaller than 0.2~eV; here, due to improvement of hydrogenation method, the gap evolution rises by 50\%.

\begin{figure}
\centering
\includegraphics[width=0.95\linewidth]{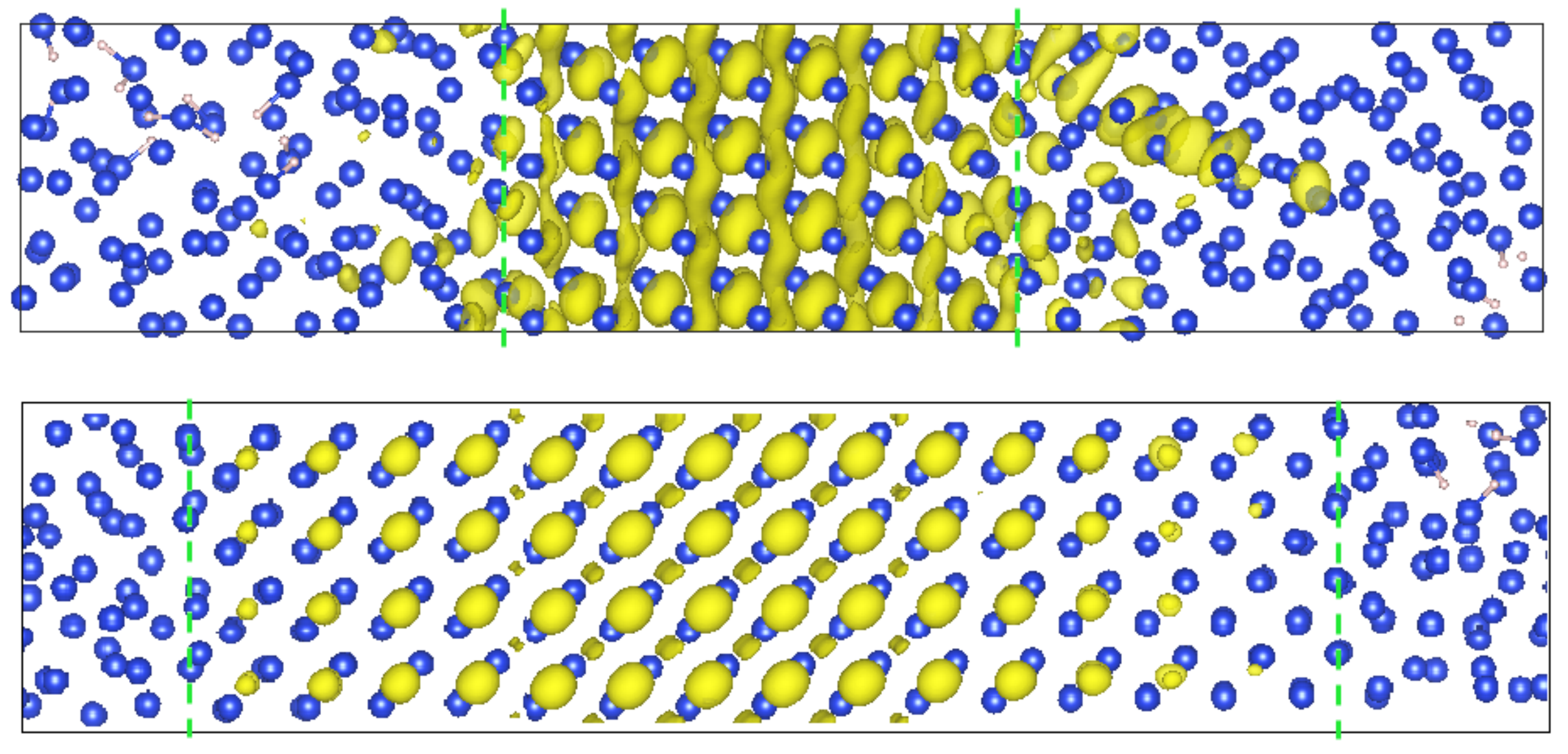}
\caption{\label{qs320} Localization of the HOMO states in the crystalline part of nc-Si:H system.
 In the shown structures, the widths of SiNSs are 1.63~nm and 4.09~nm. The isosurface level is $0.1$~e/\AA$^3$. The small pink balls in the figure are H atoms.}
\end{figure}

Figure~\ref{qs320} shows the distribution of the wave function of the HOMO states in two confined systems with the Si slab width of  1.63~nm and 4.09~nm. It shows that the HOMO states are localized in the crystalline part. This figure also reveals that even a-Si matrix as tiny as 1.36~nm is able to confine the carriers in the middle of the crystal NS. This is an important result which shows confinement power of a-Si matrix.


\begin{figure}
\centering
\includegraphics[width=0.6\linewidth]{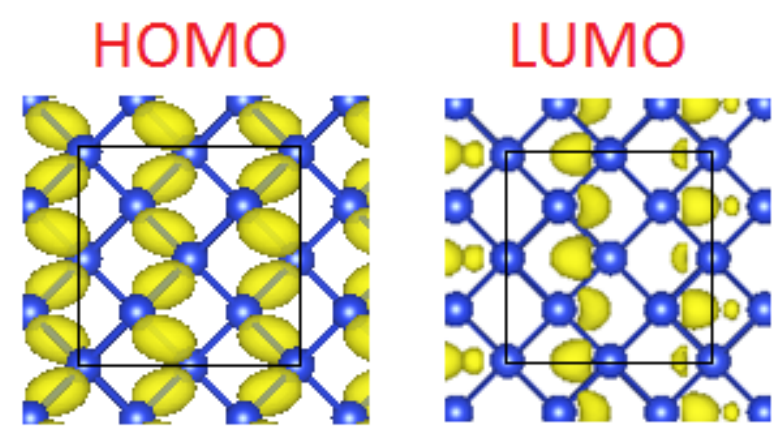}
\caption{\label{hl-state} The distribution of the valence band maximum or HOMO and the conduction band minimum or LUMO state in silicon crystal.}
\end{figure}

Figure.~\ref{gap} also concludes that in this system QC happens just for one type of carriers and electrons do not feel barriers as much as holes; consequently, the gap evolution mainly comes from confinement of holes in  the SiNSs. This notable property has been seen previously in similar systems where Si nanocrystals are confined within a-Si matrix in one, two or three directions \cite{mattoni, lusk1, lusk2}. The reason of this special kind of confinement will be discussed in this article.
To explain it, note that under normal operating conditions, silicon atoms crystallize into a tetrahedrally coordinated diamond lattice with $sp^3$ orbital hybridization; and the HOMO (LUMO) state comes from bonding (anti-bonding) between $sp^3$ orbitals \cite{SiBand}. The charge distribution for the bonding state with the symmetric space wave function is strong in the space between two atoms; contrary to this for anti-bonding orbitals with anti-symmetric space wave function, there is almost no charge density in the space between cores. As a result of that, in silicon the valence band or HOMO state, unlike the conduction band or LUMO state, is very sensitive to the bond stretching  and bond bending.  
Figure~\ref{hl-state} shows the distribution of the HOMO and the LUMO states in silicon crystal. As expected, the charge density of the HOMO state is very strong in the space between cores, while there is no charge distribution in the LUMO state. Consequently, in nc-Si:H device, confinement of HOMO states or holes is stronger than electron confinement.

\section{Summary} \label{sec:summary}

On the basis of DFT computational simulations and  employing charge density analysis techniques based on Bader's \tql atoms in molecules\tqr~ theory, in this manuscript, we study the bonding properties of a-Si system; this is the first study of this kind where as discussed in Sec. \ref{sec:bond}, its results are reflected in both mechanical and electronic properties of system.
In addition, the role of amorphous matrix on quantum confinement of charge carriers in SiNSs is discussed. It is shown that confinement effects for SiNSs limited in  [100] direction are stronger than other directions, so we consider our crystalline-amorphous Si heterojunction systems for SiNSs limited in [100] direction. Because of existence the defect states in a-Si, QC just happens in the hydrogenated systems. We used Allan formula \cite{allan} to identify defect states of a-Si, and removing the Si atoms caused these defects. The bonding properties of the removed Si atoms are also determined and discussed in Sec.~\ref{sec:bond}.
The rate of hydrogenation is found an important factor to improve the quality of a-Si electronic. 
The quantum confinement is appeared in the encapsulated SiNSs for the slab width more than $d_c=$ 1.5~nm. The power law behavior of gap tunablilty with the slab width has been extracted. Also the results imply that even a-Si  matrix as tiny as $\sim 1.3$~nm is able to confine the carriers in crystalline part. This is a promising finding for solar cell applications as the charge trapping and LID in a-Si will enhance by increasing the matrix size; in addition, it is important to reduce the silicon consumption. Finally, we explain why the confinement of holes in Si nano structures embedded in a-Si matrix are stronger than confinement of electrons.

\section*{Acknowledgments}
	
The numerical simulations are performed using the computational facilities of the School of Nano Science of IPM. Z.N. acknowledges Prof. Mark T. Lusk (CSM), Prof. Reza Asgari (IPM), and Dr. M. Goli (IPM) for valuable discussions.


\begin{thebibliography}{7}

\bibitem{qc-app}
M. D. Archer and A. J. Nozik,
\textit{Nanostructured and Photoelectrochemical Systems for Solar Photon Conversion}
(Imperial College Press, London, 2008).

\bibitem{amoSi}
R. A. Street,  
\textit{Hydrogenated Amorphous Silicon}
(Cambridge University Press, England, 2005).

\bibitem{shq}
W. Shockley and H. J.Queisser,
Appl. Phys. Lett. {\bf 32}, 510 (1961).

\bibitem{qc}
H. Haug and S. W. Koch,
\textit{Quantum Theory of the Optical and Electronic Properties of Semiconductors}
(World Scientific, Singapore, 2009).

\bibitem{meg}
M. C. Beard, K. P. Knutsen, P. Yu, J. M. Luther, Q. Song, W. K. Metzger, R. J. Ellingson, and A.J. Nozik,
Nano Lett. {\bf 7}, 2506 (2007).

\bibitem{oxide}
S. Godefroo, M. Hayne, M. Jivanescu, A. Stesmans, M. Zacharias, O. I. Lebedev, G. V. Tendeloo, and V. V. Moshchalkov,
Nature Nanotech {\bf 3}, 174 (2008).

\bibitem{nitride}
T.-Y. Kim, N.-M. Park, K.-H. Kim, G.Y. Sung, Y.-W. Ok, T.-Y. Seong, and C.-J. Choi,
Appl. Phys. Lett. {\bf 85}, 5355 (2004).

\bibitem{mattoni}
L. Bagolini, A. Mattoni, G. Fugallo, L. Colombo, E. Poliani, S. Sanguinetti, and E. Grilli,
Phys. Rev. Lett.  {\bf 104}, 176803 (2010).

\bibitem{mattoni1}
L. Bagolini, A. Mattoni, and L. Colombo,
Appl. Phys. Lett. {\bf 94}, 053115 (2009).

\bibitem{mattoni2}
G. Fugallo and A. Mattoni,
Phys. Rev. B {\bf 89}, 045301 (2014).

\bibitem{mattoni3}
S.Pizzini, M. Acciarri, S. Binetti, D. Cavalcoli, A. Cavallini, D.Chrastin, L. Colombo, E. Grilli, G. Isell, M.Lancin, A. Le Donne,  A. Mattoni, K. Peter, B. Pichaud, E. Poliani, M. Rossi, S. Sanguinetti, M. Texier, and H. von Känel,
Materials Science and Engineering: B {\bf 134}, 118 (2006).

\bibitem{lusk1}
M. T. Lusk, R. T. Collins, Z. Nourbakhsh, H. Akbarzadeh,
Phys. Rev. B {\bf 89}, 075433 (2014).

\bibitem{lusk2}
L. Bagolini, A. Mattoni, R. T. Collins, and M. T. Lusk,
J. Phys. Chem. C {\bf 118}, 13417 (2014).

\bibitem{tandem}
M. Green,
\textit{Third Generation Photovoltaics Advanced Solar Energy Conversion}
(Springer, New York, 2003).

\bibitem{lid}
J. Meier, R. Fluckiger, H. Keppner, and A. Shah,
Appl. Phys. Lett. {\bf 65}, 860 (1994).

\bibitem{k-sh}
W. Kohn and L. J. Sham,
Phy. Rev. {\bf 140}, A1133 (1965).

\bibitem{bader}
R. Bader, \textit{Atoms in Molecules: A Quantum Theory}
(Oxford University Press, UK, 1994).

\bibitem{bader1}
H. Liu, J. S. Tse, M. Y. Hu, W. Bi, J. Zhao, E. E. Alp, M. Pasternak, R. D. Taylor, and J. C. Lashley,
J. Chem. Phys. {\bf 143}, 164508 (2015).

\bibitem{direction}
T. E. Jones, M. E. Eberhart, and D. P. Clougherty,
Phys. Rev. Lett. {\bf100}, 017208 (2008).

\bibitem{nour}
Z. Nourbakhsh and R. Asgari,
Phys. Rev. B {\bf 98}, 125427 (2018).

\bibitem{bulkm1}
Z. Nourbakhsh, S. J. Hashemifar, H. Akbarzadeh,
J. Alloy. Compd. {\bf579}, 360 (2013).

\bibitem{bulkm2}
Z. Nourbakhsh, S. J. Hashemifar, H. Akbarzadeh,
J. Magn. Magn. Mater. {\bf 341}, 56 (2013). 

\bibitem{bader3}
J. Miorelli, T. Wilson, A. Morgenstern, T. Jones, and M. E. Eberhart,
ChemPhysChem {\bf 16}, 152 (2015). 

\bibitem{bader4}
Y. Yue, Y. Song, and X. Zuo,
AIP Advances {\bf 7}, 015309 (2017).
L. Villamagua, R. Rivera, D. Castillo, and M. Carini,
AIP Advances {\bf 7}, 105010 (2017).

\bibitem{www}
F. Wooten, K. Winer, and D. Weaire,
Phys. Rev. Lett. {\bf 54}, 1392 (1985).

\bibitem{cleangap}
G. T. Barkema and N. Mousseau,
Phys. Rev. B {\bf 62}, 4985 (2000).

\bibitem{exp}
K. Laaziri, S. Kycia, S. Roorda, M. Chicoine, J. L. Robertson, J. Wang, and S.C. Moss,
Phys. Rev. B {\bf 60}, 13520 (1999).

\bibitem{siesta}
J. M. Soler, E. Artacho, J. D. Gale, A. Garcia, J. Junquera, P. Ordejon, D. Sanchez-Portal,
J. Phys.: Condens.Matter {\bf14}, 2745 (2002).
http://departments.icmab.es/leem/siesta/

\bibitem{pbe}
J. P. Perdew, K. Burke, and M. Ernzerhof,
Phys. Rev. Lett. {\bf 77}, 3865 (1996).

\bibitem{aim}
http://theory.cm.utexas.edu/henkelman/code/bader/

\bibitem{allan}
G. Allan, C. Delerue, and M. Lannoo,
Phy. Rev. B {\bf 57}, 6933 (1998).

\bibitem{hsol}
M. Stavola,
\textit{H Diffusion and Solubility in c-Si}
(INSPEC, London, 1999).

\bibitem{expgap}
G. Kaniadakis,
\textit{Properties of Amorphous Silicon}
(INSPEC, London, 1989).

\bibitem{aSiMech}
C. L. Allred, X. Yuan, M. Z. Bazant, and L. W. Hobbs,
Phys. Rev. B {\bf70}, 134113 (2004).

\bibitem{aSiMech2}
J. D. Schall, G. Gao, and J. A. Harrison,
Phys. Rev. B {\bf 77}, 115209 (2008).

\bibitem{grosso}
G. Grosso and G. P. Parravicini,
\textit{Solid State Physics}
(Academic Press, London, 2000).

\bibitem{dftgap3}
M. Gr{\"u}ning,  A. Marini, and A. Rubio,
J. Chem. Phys.  {\bf 124}, 154108 (2006).

\bibitem{SiBand}
W. A. Harrison,
\textit{Elementary Electronic Structure}
(World Scientific, Singapore, 1999); 
W. A. Harrison,
\textit{Electronic Structure and the Properties of Solids: The Physics of the Chemical Bond}
(Dover Publications, Inc, New York, 1989).

\end{thebibliography}
\end {document}